\documentstyle[preprint,aps,prl,eqsecnum]{revtex}
\begin{document}
\preprint{ISSP-94}

\title{Excitation Spectrum of $S=1$ Antiferromagnetic Chains }
\author{Minoru Takahashi }
\address{Institute for Solid State Physics, University of Tokyo}
\address{Roppongi, Minato-ku, Tokyo 106, Japan}
\date{12-04-1994}
\maketitle
\begin{abstract}
The dynamical structure factor $S(Q,\omega)$ of the $S=1$ 
antiferromagnetic Heisenberg chain with length 20 at zero temperature 
is calculated. The lowest energy states have the 
delta-function peak at the region $\pi\ge \vert Q\vert >0.3\pi$. 
At $\vert Q\vert<0.3\pi$ the lowest energy states are the lower-edge of the 
continuum of the scattering state, the strength of which 
decreases for large systems. This gives a reasonable 
explanation for the experimental fact that no clear peak is observed 
at the region $Q<0.3\pi$. This situation is more apparent 
for valence-bond solid state. On the contrary for $S=1/2$ antiferromagnetic 
Heisenberg chain the lowest energy states are always the edge of 
the continuum.    
\end{abstract}
\newpage
\section{INTRODUCTION}
\label{sec:intro}
Haldane~\cite{haldane} first argued that integer-spin $S$ 
antiferromagnetic Heisenberg chains have a singlet ground state 
which has a gap to a triplet excited states.
This has now been well confirmed both experimentally~\cite{expgap,renard87}
and numerically~\cite{numgap,tak1,white1,white2,tak93,nom89} for $S=1$ system.
Recently, Ma et al carried out detailed inelastic neutron scattering 
experiments~\cite{ma} on Ni(C$_2$H$_8$N$_2$)$_2$NO$_2$(ClO$_4$) (NENP), 
which is one of the most promising candidates for a Haldane gap system. 
They measured $S(Q,\omega )$ from $Q=\pi$ down to $Q=0.3 \pi$ where
the intensity became weak.  At all $Q$, the peak width
is the resolution limit.  Their experimental results coincide 
very well with numerical results of static correlation functions 
and low-lying excitations. The low-lying state 
gives the delta-function type peak in the region 
$\pi\ge \vert Q\vert >0.3\pi$. It was stated that the gap near 
$Q=0$ is twice of that near $Q=\pi$~\cite{tak1}. 
This means that the lowest energy 
state near $Q=0$ is made up of two excitations near $Q=\pi$. Thus, 
near $Q=0$ the 
lowest energy state should be the lower edge of the continuum made by 
the scattering states. On the contrary, however, 
near $Q=\pi$ the lowest energy 
state  has a strong delta-function peak. There must  
,then, be a change in the property of the peak at a certain momentum. 
The Monte Carlo method is not appropriate to use to investigate the details of 
$S(Q,\omega)$, so we used the diagonalization method and attempted to treat 
longer systems. This allows calculation 
not only of the first excited state but also of higher excited states and  
their contribution to the dynamical structure factor $S(Q,\omega)$~\cite{gb}. 
The Hamiltonian we investigate in this paper is as follows:
$${\cal H}=\sum_{i=1}^N J[S_i^xS_{i+1}^x+S_i^yS_{i+1}^y+\lambda S_i^zS_{i+1}^z
-\beta({\bf S}_i\cdot{\bf S}_{i+1})^2]+D(S_i^z)^2,$$
$${\bf S}_i^2=2,\quad {\bf S}_{N+1}={\bf S}_1.\eqno(1)$$
The real NENP system is well described by $\lambda=1, D=0.18J$ and $\beta=0$.
~\cite{gol}
In a previous paper~\cite{tak93} we calculated the cases 
$\lambda=1, D=0,\pm 0.2J$ 
and  $\beta=0$ up to $N=18$, and found that 
most of the spectral weight is concentrated in the lowest energy state. 
The ratio is more than 90\% except near $Q=0$. 
We then wanted to investigate what happens 
at small momenta and scattering intensity of higher excited states.  
The ground state was investigated by several authors~\cite{aklt,aah} 
about the point $\lambda=1, D=0, \beta=-1/3$ . 
Following Affleck, Kennedy, Lieb and 
Tasaki~\cite{aklt} we call this point the valence-bond solid (VBS) point,  
and we call the point $\lambda=1, D=0, \beta=0$ 
the antiferromagnetic Heisenberg (AFH) point. In this paper 
we carefully investigate the excitation spectra at these 
two points. \par
We succeeded in  
diagonalizing $N=20$ chain and determining the dynamical structure factor. 
The scattering intensity of the low-lying states 
decreases considerably in the region 
$Q<0.3\pi$. Moreover, near $0.3\pi$ the second excited state approaches 
the first excited state. These facts mean that the single 
delta-function 
peak disappears at the region $Q<Q_c$. This coincides with the 
experimental 
fact that Ma could not observe any clear peak in this momentum region. 
For comparison we calculate the dynamical structure factor at the VBS 
point ($\beta=-1/3$). Qualitatively the situation is very similar to 
the AFH case ($\beta=0$). 
In VBS case the single delta-function peak is absorbed in 
the continuum at $Q=0.4\pi$. The lowest energy state becomes the lower edge 
of $S(Q,\omega)$ and the intensity drops more rapidly and clearly.  \par
\S 2A summerizes our new method of representing states in the 
subspace of momentum $Q$. In our representation the Hamiltonian is 
always a real and symmetric matrix. State vectors 
have real elements. In the conventional representation Hamiltonian 
is hermitian and state vectors have complex elements. 
The memory necessary in our method is about half that 
of the conventional method.  We can 
diagonalize $N=20$ systems. \par
\S 2B summarizes the method of Gagliano and Balseiro 
which is useful to calculate the dynamical structure factor. \par
\S3 compares the $S=1$ antiferromagnetic Heisenberg (AFH) chain, $S=1$ 
VBS chain and $S=1/2$ AFH chain. We find that the 
$S=1$ AFH chain is qualitatively 
very near the $S=1$ VBS chain, but the $S=1/2$ AFH chain is completely 
different. We establish that the lowest energy state at $Q>Q_c$ has 
macroscopic strength and that at $Q<Q_c$ the lowest energy state 
is merely the lower edge of the continuum. The strength becomes microscopic.  
The lowest energy states of $S=1/2$ AFH chain, on the other hand,
are always the lower edge of the continuum and the strength 
is always microscopic.    
 
\section{NUMERICAL METHOD}
\label{sec:nume}
\subsection{Real symmetric matrix in momentum subspace}
As each site has three states, the Hamiltonian (1) is a matrix in $3^N$ 
dimensional space. Spin configuration of a state is represented by a 
trinary number with length $N$ as 
$$\vert -1,0,0,-1,1,0>=\vert 011021_3>.\eqno(2)$$
We introduce the shift operator ${\bf T}$ and inversion operator ${\bf R}$ 
as follows:
$${\bf T}\vert 011021_3>=\vert 110210_3>,\quad {\bf T}^N=I,$$
$${\bf R}\vert 011021_3>=\vert 120110_3>,\quad{\bf R}^2=I,\quad  
{\bf R}{\bf T}^l={\bf T}^{N-l}{\bf R}.\eqno(3)$$
Both operators commute with the Hamiltonian (1). 
Here $I$ is the identity operator. 
We can classify the subspace 
by $S^z_{total}$ and momentum $Q=2\pi \times integer/N$. 
Each subspace is 
expressed by the following set of bases:
$$\vert a,Q>=const. \sum_{l=0}^{N-1}\exp(-iQl){\bf T}^l\vert 011021_3>,
\quad {\bf T}\vert a,Q>=\exp(iQ)\vert a,Q>.\eqno(4)$$
We use the smallest trinary numbers in ${\bf T}^l \vert$trinary number$>$ 
to represent a state $\vert a,Q>$.  
In this representation the Hamiltonian is a Hermite matrix 
which has the complex matrix elements, except the cases $Q=0,$ or $\pi$. 
The elements of eigenvectors are also 
complex. For $N=20, S^z_{total}=0$ case, the dimensions of the subspace 
are $1.89\times10^7$. 
If we can represent the Hamiltonian as a real symmetric 
matrix the memory and computing time should be reduced to half for the 
Hermite matrix. 
To find the real representation of Hamiltonian we classify the bases as 
follows. Let us call a base $\vert a,Q>$ symmetric if 
$$R\vert a,Q>=\exp(-iQ l_a){\vert a,-Q>},\eqno(5a)$$ 
is satisfied for certain 
integer $l_a$, and asymmetric if not. An asymmetric base $\vert b,Q>$ 
must have its counterpart $\vert  \overline b,Q>$ which satisfies:
$$ R\vert b,Q>=\exp(-iQ l_b){\vert  \overline b,-Q>},\quad 
R\vert  \overline b,Q>=\exp(-iQ l_b){\vert b,-Q>}.\eqno(5b)$$
For example, the state $(011011)$ is symmetric and $l_a=1$. The state 
$(011012)$ is asymmetric and $(011021)$ is its counterpart and $l_b=4$.     
We generate a new set of bases as follows:
$$\vert a',Q>=\exp(iQ l_a/2)\vert a,Q>,$$
$$\vert b',Q>=2^{-1/2}\exp(iQl_b/2)( \vert b,Q>+\vert \overline b,Q>),$$
$$\vert \overline b',Q>=-i 2^{-1/2}\exp(iQl_b/2)
( \vert b,Q>-\vert \overline b,Q>).\eqno(6)$$
In this new set of bases the Hamiltonian matrix is real and symmetric. 
Each column of the Hamiltonian has $2N$ non-zero off-diagonal elements 
at most. We therefore calculate the values and positions of non-zero 
off-diagonal elements and store them in an 
auxiliary memory before the Lanczos calculation. Thus only 
three vectors are needed in the core memory.
 
\subsection{Gagliano-Balseiro method}
We first calculate the lowest energy state $\vert 0>$ in the subspace 
$S^z_{total}=0, Q=0$ , namely the ground state using the Lanczos method. 
In the subspace $S^z_{total}=0, Q\ne 0$ we use 
$S_Q^z\vert 0>$ as the initial vector of the Lanczos calculation. 
Following Gagliano and Balseiro~\cite{gb} we calculate 
the dynamical structure factors 
$$S_{\parallel}(Q,\omega)=
\sum_n \vert <n\vert S_Q^z\vert 0>\vert^2\delta[\omega-(E_n-E_0)/\hbar],$$
$$S_{\perp}(Q,\omega)=
\sum_n \vert <n\vert S_Q^-\vert 0>\vert^2
\delta[\omega-(E_n-E_0)/\hbar].\eqno(7)$$
$S_{\parallel}(Q,\omega)$ is given by the Green function:
$$S_{\parallel}(Q,\omega)=-{\hbar\over \pi}
{\rm Im}G_{\parallel}(Q,\hbar\omega),$$  
$$G_{\parallel}(Q,z)\equiv 
<0\vert S_{-Q}^z(z-{\cal H})^{-1}S_Q^z\vert 0>.\eqno(8)$$ 
For any operator $A$ the Green function is given as a continued fraction by 
the Laczos method.
$$<0\vert A^{\dagger}{1\over z-{\cal H}}A\vert 0>
={<0\vert A^{\dagger}A\vert 0>
\over{z-a_1-{b_1^2\over{z-a_2-{b_2^2\over z-a_3-...}}}}},\eqno(9)$$
$$\vert f_0>=A\vert 0>,\quad b_0=0,$$
$$\vert f_{n+1}>={\cal H}\vert f_{n}>-a_n\vert f_{n}>
-b_n^2\vert f_{n-1}>,$$
$$a_n=<f_n\vert {\cal H}\vert f_n>/<f_n\vert f_n>,$$
$$b_{n+1}^2=<f_{n+1}\vert f_{n+1}>/<f_n\vert f_n>.\eqno(10)$$
We can determine poles and residues of this continued fraction 
and obtain the scattering intensities of each state. 
In the same way we can calculate $S_{\perp}(Q,\omega)$. 
For high energy states this method does not give accurate results, 
but for low-lying states works very well~\cite{haas}.  
 
\par
\section{RESULTS FOR SPIN CHAINS}
\label{sec:res}
\subsection{Isotropic Heisenberg chain}

In Fig.1 we plot the scattering intensities of the lowest energy states 
of each momentum at isotropic Heisenberg point. At $Q>0.3\pi$ 
the ratio of 
scattering intensity of lowest energy state is more than 90\% and not 
dependent 
on the size $N$. At $Q<0.3\pi$ the ratio decreases considerably as 
$N$ becomes large. The 
size dependence is strong and intensity seems to go to zero as $N$ goes to 
infinity. In Fig.2 we plot the first and second poles of momentum $Q$. 
We find that two states have almost the same energy at $Q=0.3\pi$, 
this is also evidence that the properties of the low-lying states 
change from this momentum.  Figure 3 is the static structure factor $S(Q)$. 
Figure 4 is the poles and residues of the dynamical structure factor 
$S(Q,\omega)$. There is considerable distance between the first 
and the second excited state at $Q>0.3\pi$.
\subsection{VBS chain}
As the correlation length $\xi=6.1$ is comparable with size $N=20$ 
the decrease in intensity of the lowest energy state is not clear for the 
AFH chain. 
But this situation becomes more clear if we calculate the ratio of the 
lowest energy state for the VBS chain. As shown in Fig.5 we find that the ratio 
goes to zero at $Q<0.4\pi$. In the VBS case the correlation length 
$\xi(=1/\ln 3)$ is much smaller than the system size and there is a clear 
decrease in the scattering intensity of the lowest energy state.
In Fig.6 the first and the second excited energies are shown.  
Arovas, Auerbach and Haldane~\cite{aah} obtained the structure 
factor $S(Q)$ at the VBS point. Our result of $S(Q)$ in Fig.7 
reproduces their result. 
$$<0\vert S_{-Q}^zS_Q^z\vert 0>={2(1-\cos Q)\over 5+3\cos Q}.\eqno(11)$$
Then, the variational energy becomes:
$$\omega_Q={1\over2}
{<0\vert [S_{-Q}^z,[{\cal H},S_Q^z]]\vert 0>\over 
<0\vert S_{-Q}^zS_Q^z\vert 0>}={10\over 27}(5+3\cos Q).\eqno(12)$$
The lowest energy state is found very near this variational 
energy at $0.4\pi\le Q\le \pi$, but it enters into the continuum at 
$Q<0.4\pi$. In this region $S(Q,\omega)$ has only a vague peak 
near the variational energy, while the lowest energy state is far from 
this energy. There is considerable distance between the first 
and the second excited state at $Q>0.4\pi$.

\subsection{$S=1/2$ Heisenberg antiferromagnet }

It is known that the $S=1/2$ Heisenberg model (
${\cal H}=J\sum {\bf S}_i\cdot{\bf S}_{i+1},\quad {\bf S}_i^2=3/4$) 
has the low-lying excitation 
$\epsilon(Q)={\pi\over 2}J\sin\vert Q\vert$. 
We calculate the ratio of the scattering intensity of this 
excitation in the states of momentum $Q$. 
As shown in Fig.9. they decrease as $N$ becomes large. 
Then, for this system the lowest energy state at fixed momentum 
is the bottom of the continuum made by scattering states of 
two excitations with spin 1/2 as was noted 
by Faddeev and Takhtajan~\cite{fadtak}. This continuum 
has been justified experimentally~\cite{tenn}.
The strength of each excitation is microscopic. 
The  distance between the first and the second excitations decreases as 
the system enlarges. In Fig.11 we show the structure factor of this system. 
It has a logarithmic peak at $Q=\pi$. In Fig.12 we show the dynamical 
structure factor of $N=24$ $S=1/2$ AFH chain.    \par
\section{CONCLUSION AND DISCUSSION}
From these numerical results we can conclude the following for 
$S=1$ Haldane antiferromagnets: \par\noindent 
1)At $\pi\ge Q >Q_c$, the lowest energy state gives 
a scattering intensity of 
more than 90\% of total intensity of 
the states of momentum $Q$. At $Q<Q_c$ 
the lowest energy state is the lower edge of the continuum and 
does not have the delta-function type peak. \par\noindent
2)The value of $Q_c$ is about $0.3\pi$ for isotropic 
Heisenberg point and about $0.4\pi$ for VBS point.    \par
It is very interesting that the kink in 
$\epsilon(Q)$ at $Q=Q_c$ and $\beta=0$ is too weak to be observed. \par
The author is grateful to Ian Affleck 
and Stephen Haas for stimulating discussions. 
Most numerical calculations were done by HITAC S-3800 at the 
computer center of the University of Tokyo. 
This work was supported in part by 
Grants-in-Aid for Scientific Research on 
Priority Areas, "Computational Physics as a New Frontier in Condensed 
Matter Research" (Area No 217) and "Molecular Magnetism" (Area No 228), 
from the Ministry of Education, Science and Culture, Japan. 
\appendix \section*{Proof of real-symmetriness of Hamiltonian}
We show that Hamiltonian is real and symmetric in the bases given 
in equation (6). It is sufficient if we prove the symmetriness, because 
the Hamiltonian matrix is Hermitian.  
We first consider matrix elements between two symmetric 
states $\vert a_1',Q>$ and $\vert a_2',Q>$: 
$$<a_2',Q\vert{\cal H}\vert a_1',Q>=
\exp(iQ(l_1-l_2)/2)<a_2,Q\vert{\cal H}\vert a_1,Q>.\eqno(A1)$$
Using Eq.(4a) and ${\cal H}={\bf R}{\cal H}{\bf R}$ we have
$$(A1)=\exp(-iQ(l_1-l_2)/2)<a_2,-Q\vert{\cal H}\vert a_1,-Q>.\eqno(A2)$$
As the Hamiltonian is Hermitian in conventional representation we have:
$$(A2)=\exp(-iQ(l_1-l_2)/2)\overline{<a_1,-Q\vert{\cal H}\vert a_2,-Q>}$$
$$=\exp(-iQ(l_1-l_2)/2)<a_1,Q\vert{\cal H}\vert a_2,Q>
=<a_1',Q\vert{\cal H}\vert a_2',Q>.\eqno(A3)$$
Thus we know that $<a_2',Q\vert{\cal H}\vert a_1',Q>$ and 
$<a_1',Q\vert{\cal H}\vert a_2',Q>$ is the same and is real. \par
Next we consider elements between symmetric and asymmetric states:
$$<b',Q\vert {\cal H}\vert a',Q>=\exp(iQ(l_a-l_b)/2)2^{-1/2}
\{<b,Q\vert +<\overline{b},Q\vert\}{\cal H}\vert a,Q>.\eqno(A4)$$
Using Eq.(4b) and ${\cal H}={\bf R}{\cal H}{\bf R}$ we have
$$(A4)=2^{-1/2}\exp(-iQ(l_a-l_b)/2)
\{<b,-Q\vert +<\overline{b},-Q\vert\}{\cal H}\vert a,-Q>.\eqno(A5)$$
By Hermiticity we have:
$$(A5)=2^{-1/2}\exp(-iQ(l_a-l_b)/2)\overline{<a,-Q\vert{\cal H}
\{\vert b,-Q>+\vert \overline{b},-Q>\} }$$
$$=2^{-1/2}\exp(-iQ(l_a-l_b)/2)<a,Q\vert{\cal H}
\{\vert b,Q>+\vert \overline{b},Q>\}=<a',Q\vert{\cal H}\vert b',Q>.\eqno(A6)$$
In the same way we can prove the symmetriness of elements between 
$\vert \overline{b}',Q>$ and $\vert a',Q>$. \par
$$<\overline{b}',Q\vert {\cal H}\vert a',Q>=i\exp(iQ(l_a-l_b)/2)2^{-1/2}
\{<b,Q\vert -<\overline{b},Q\vert\}{\cal H}\vert a,Q>$$
$$=-i2^{-1/2}\exp(-iQ(l_a-l_b)/2)
\{<b,-Q\vert -<\overline{b},-Q\vert\}{\cal H}\vert a,-Q>$$
$$=-i2^{-1/2}\exp(-iQ(l_a-l_b)/2)\overline{<a,-Q\vert{\cal H}
\{\vert b,-Q>-\vert \overline{b},-Q>\} }$$
$$=-i2^{-1/2}\exp(-iQ(l_a-l_b)/2)<a,Q\vert{\cal H}
\{\vert b,Q>-\vert \overline{b},Q>\}
=<a',Q\vert{\cal H}\vert \overline{b}',Q>.\eqno(A7)$$
\par
In the last stage we must prove the symmetriness of elements between states $b_1'$,$\overline{b}_1'$ and states $b_2'$,$\overline{b}_2'$. We have 
three cases which can be proved in the same way:
$$<b_1',Q\vert {\cal H}\vert b_2',Q>=<b_2',Q\vert {\cal H}\vert b_1',Q>,$$
$$<\overline{b}_1',Q\vert {\cal H}\vert b_2',Q>=
<b_2',Q\vert {\cal H}\vert \overline{b}_1',Q>,$$
$$<\overline{b}_1',Q\vert {\cal H}\vert \overline{b}_2',Q>
=<\overline{b}_2',Q\vert {\cal H}\vert \overline{b}_1',Q>.\eqno(A8)$$
Thus, in our representation the Hamiltonian matrix is symmetric and real for 
any momentum $Q$.
\par

\begin{figure}
\caption{Fraction of the total scattering 
intensity concentrated in the lowest energy state for $N=16,18,20$ chains 
for isotropic Heisenberg point. 
In the region $Q>0.3\pi$ the ratio is more than 90\% 
and a stable function of momentum $Q$, but it decreases considerably 
as $N$ increases in the region $Q<0.3\pi$. }
\label{fig:heir}
\end{figure}

\begin{figure}
\caption{Energy of the first and second excitations of $S=1$ 
isotropic Heisenberg point. They are very close together at $Q=0.3\pi$. }
\label{fig:hei12}
\end{figure}

\begin{figure}
\caption{Structure factor $S(Q)$ of $S=1$ 
isotropic Heisenberg model. It has a Lorentzian peak at $Q=\pi$ and 
behaves as $Q^2$ near $Q=0$. }
\label{fig:heisq}
\end{figure}

\begin{figure}
\caption{Dynamical Structure factor $S(Q,\omega)$ of $N=20,\quad S=1$ 
isotropic Heisenberg model. It is made of delta-function peaks at each 
excited state. The strength is expressed as a percentage of total 
strength at the momentum $Q$.}
\label{fig:heids}
\end{figure} 

\begin{figure}
\caption{Fraction of the total scattering 
intensity concentrated in the lowest energy state 
for $N=16,18,20$ chains for VBS point. 
In the region $Q>0.4\pi$ the ratio is more than 90\% 
and a stable function of momentum $Q$. In the region $Q<0.4\pi$ it decreases 
considerably as $N$ increases. }
\label{fig:vbsr}
\end{figure}

\begin{figure}
\caption{Energy of the first and the second excitations of $S=1$ 
VBS model. They are very close together at $Q=0.4\pi$. }
\label{fig:vbs12}
\end{figure}

\begin{figure}
\caption{Structure factor $S(Q)$ of $S=1$ 
VBS model. It has a Lorentzian peak at $Q=\pi$ and 
behaves as $Q^2$ near $Q=0$. It obeys eqn.(11) exactly. }
\label{fig:vbssq}
\end{figure}

\begin{figure}
\caption{ Dynamical structure factor $S(Q,\omega)$ of $N=20,\quad S=1$ 
VBS model. It is made of delta-function peaks at each 
excited state. The strength is expressed as percentage of total 
strength at the momentum $Q$. The delta-function peak enters into 
the continuum and changes to a round peak. Both are very close to the 
variational energy given by eqn.(12). }
\label{fig:vbsds}
\end{figure}

\begin{figure}
\caption{Fraction of the total scattering 
intensity concentrated in the lowest energy state 
of $S=1/2$ Heisenberg chain for $N=16,20,24$ chains. The size-dependence 
is strong in all areas of momenta. 
This means that the lowest energy states are the bottom of the continuum 
made by two spinon excitations.   }
\label{fig:hafr}
\end{figure}

\begin{figure}
\caption{Energy of the first and second excitations of $S=1/2$ 
isotropic Heisenberg point. It seems that there is no gap 
between them.  }
\label{fig:haf12}
\end{figure}

\begin{figure}
\caption{Structure factor $S(Q)$ of $S=1/2$ 
isotropic Heisenberg chain. It has logarithmically divergent peak 
at $Q=\pi$ and behaves as $\vert Q\vert$.}
\label{fig:hafsq}
\end{figure}
\begin{figure}
\caption{Dynamical structure factor $S(Q,\omega)$ of $N=24,\quad S=1/2$ 
isotropic Heisenberg model. It is made of delta-function peaks at each 
excited state. The strength is expressed as a percentage of total 
strength at the momentum $Q$.}
\label{fig:hafds}
\end{figure}

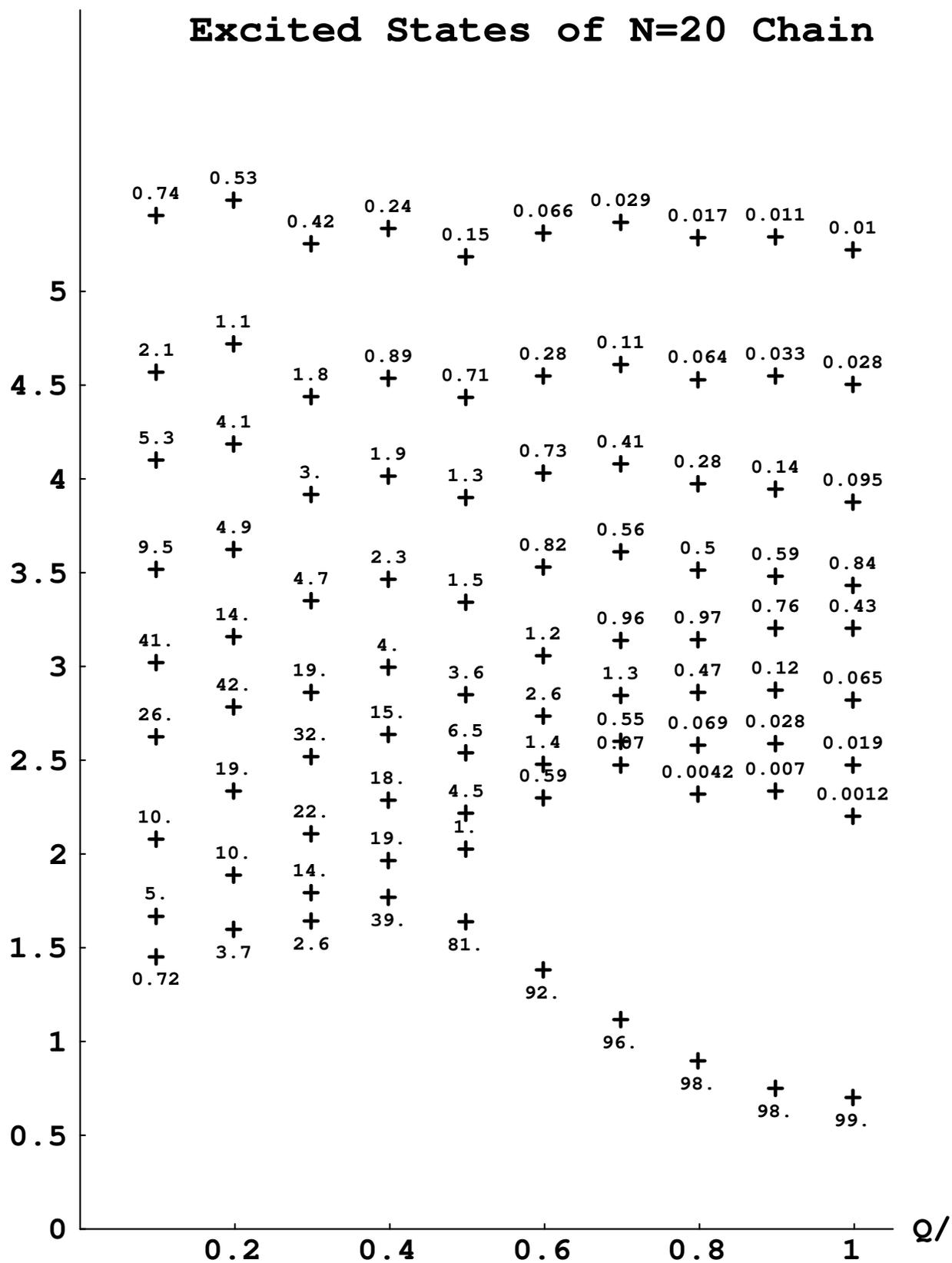
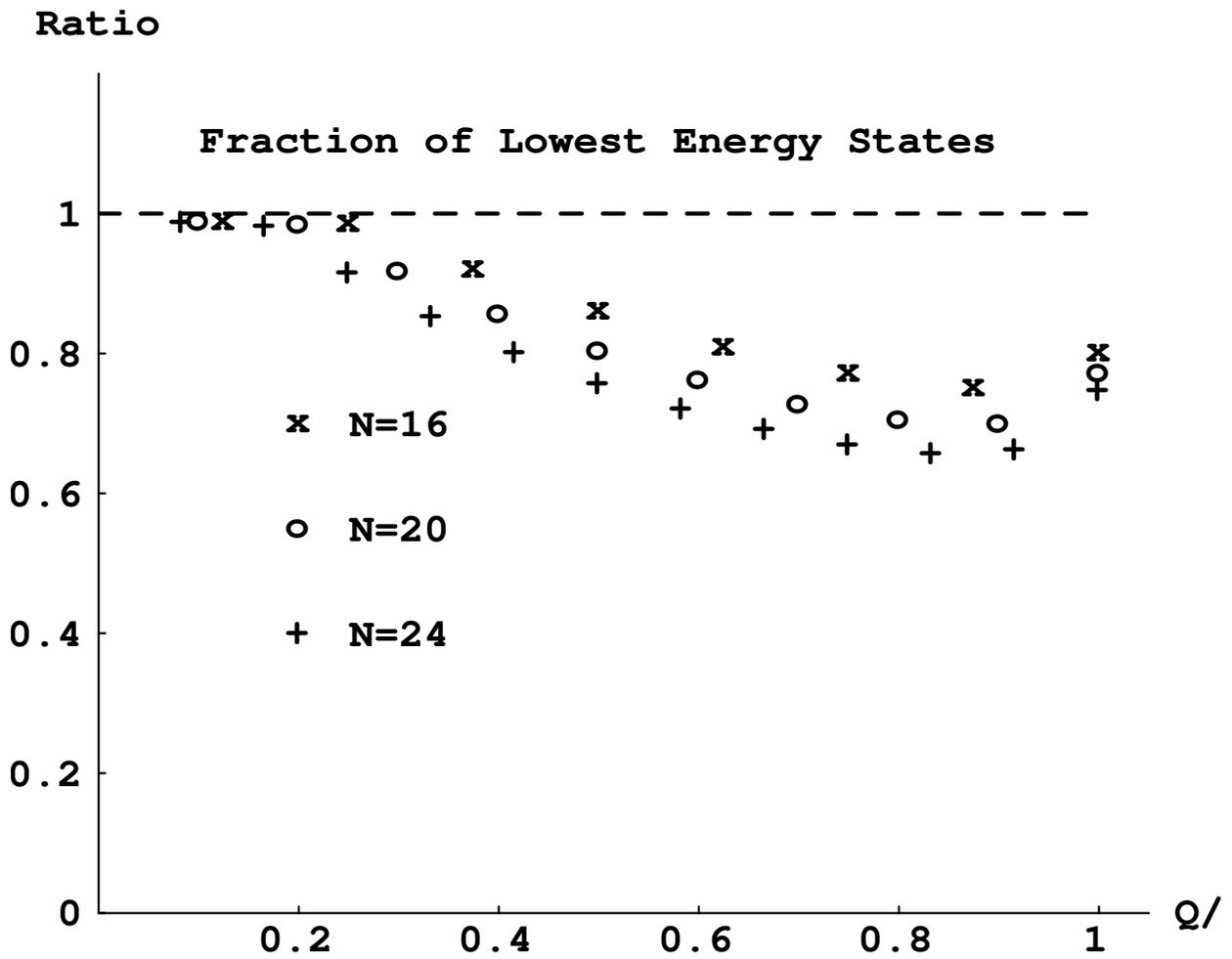
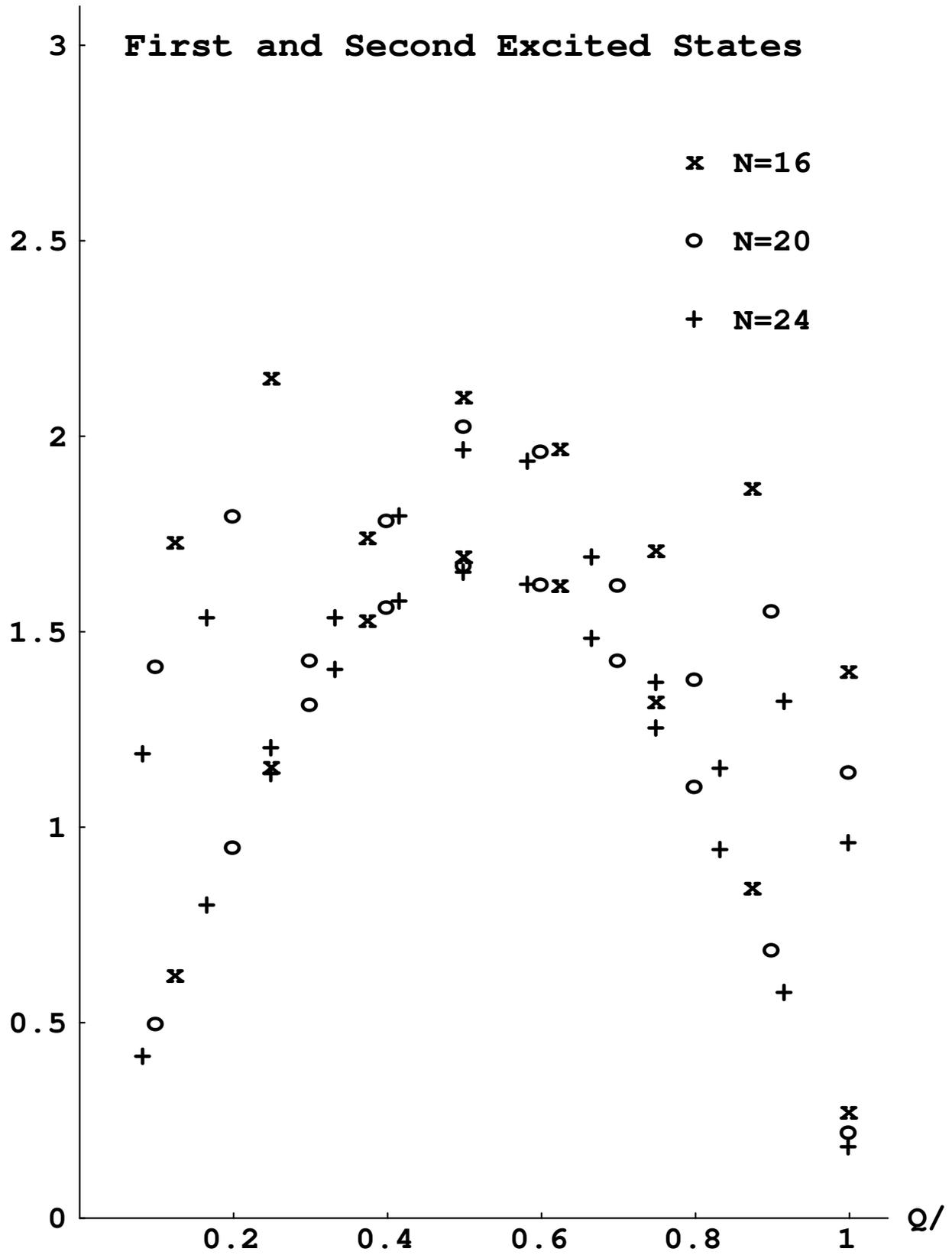
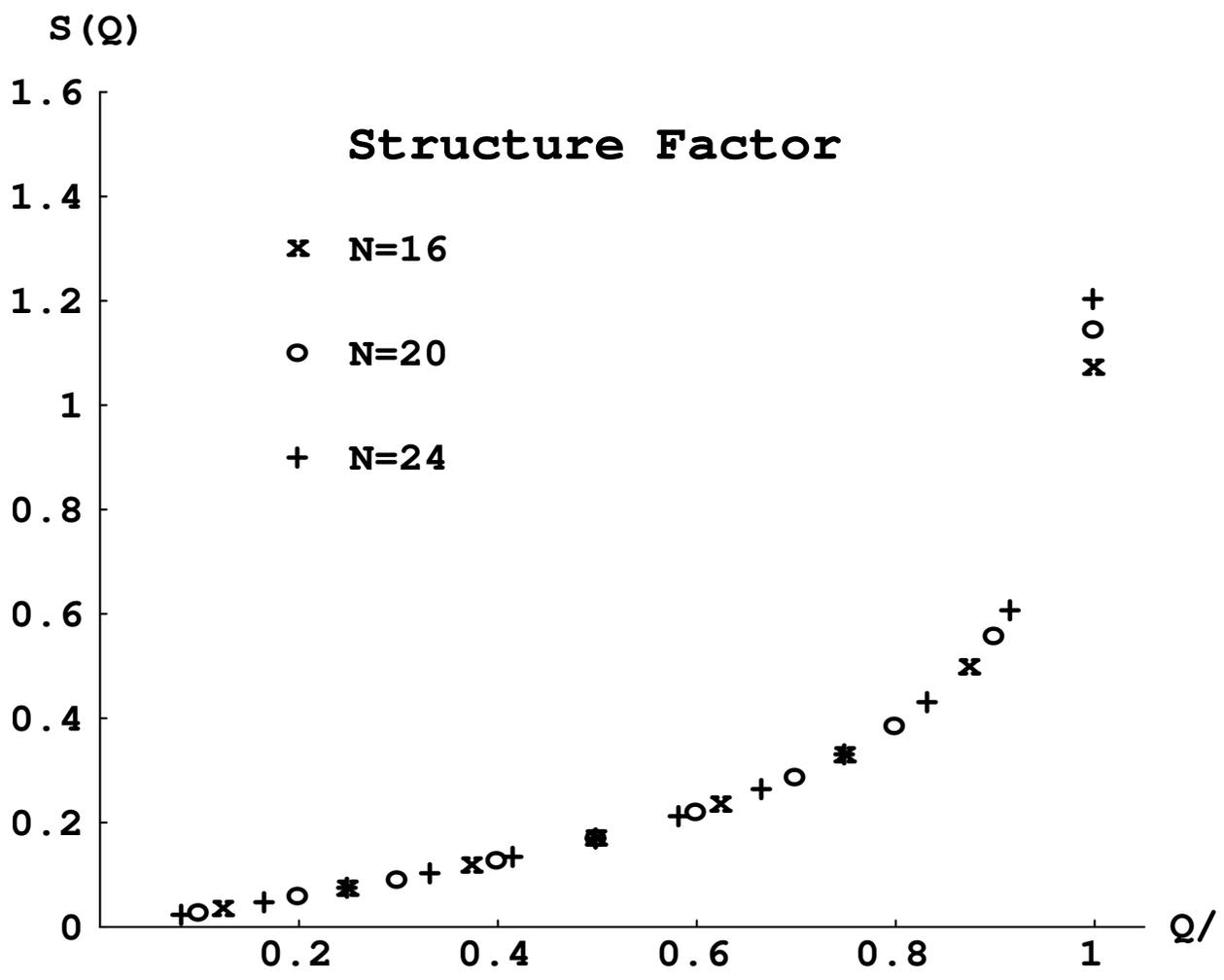
\end{document}